\documentclass[prl,twocolumn,showpacs,preprintnumbers,superscriptaddress,
amsmath,amssymb,tightenlines,epsfig]{revtex4}

\usepackage{bm}
\usepackage{graphicx}

\newcommand{\rv}{{\bf r}}

\newcommand{\dv}{{\bf d}}

\newcommand{\beq}{\begin{equation}}
\newcommand{\eeq}{\end{equation}}
\newcommand{\bea}{\begin{eqnarray}}
\newcommand{\eea}{\end{eqnarray}}

\newcommand{\<}{\langle}
\renewcommand{\>}{\rangle}

\newcommand{\commentout}[1]{{}}

\begin{document}

\title{Monopole core instability and Alice rings in spinor Bose-Einstein condensates}
\author{J.\ Ruostekoski}
\affiliation{Department of Physical Sciences, University of
Hertfordshire, Hatfield, Herts, AL10 9AB, UK}
\email{j.ruostekoski@herts.ac.uk}
\author{J.\ R.\ Anglin}
\affiliation{Center for Ultracold Atoms, Massachusetts Institute of Technology, 77 Massachusetts Ave, Cambridge MA 02139}
\email{janglin@mit.edu}

\begin{abstract}
We show how the length scale hierarchy, resulting from different interaction strengths in an optically-trapped spin-1 $^{23}$Na Bose-Einstein condensate, can lead to intriguing core deformations in singular topological defects. In particular, a point defect can be unstable with respect to the formation of a stable half-quantum vortex ring (an `Alice ring'), providing a realistic scheme to use dissipation as a sophisticated state engineering tool. We compute the threshold for stability of the point monopole, which is beyond the current experimental regime.
\end{abstract}

\date{\today}
\maketitle

The rich order parameter space of multi-component Bose-Einstein
condensates (BECs) can admit truly 3D topological excitations \cite{RUO01,SAV03,BAT02,Stoof,STO01}, beyond the simple quantized vortices of single-component BECs. Such structures are of interest in a wide range of physical contexts, but dilute atomic BECs offer the unusual advantage that we can fully explore, e.g., the short-range physics in topological defect cores, where the order parameter may explore a larger space than the usual ground state manifold. In this Letter we show how this can result in rich and surprising core structures, by demonstrating a spontaneous deformation of a singular point defect to an energetically stable half-quantum vortex ring.

\begin{figure}[tbp]
\includegraphics[width=0.5\columnwidth]{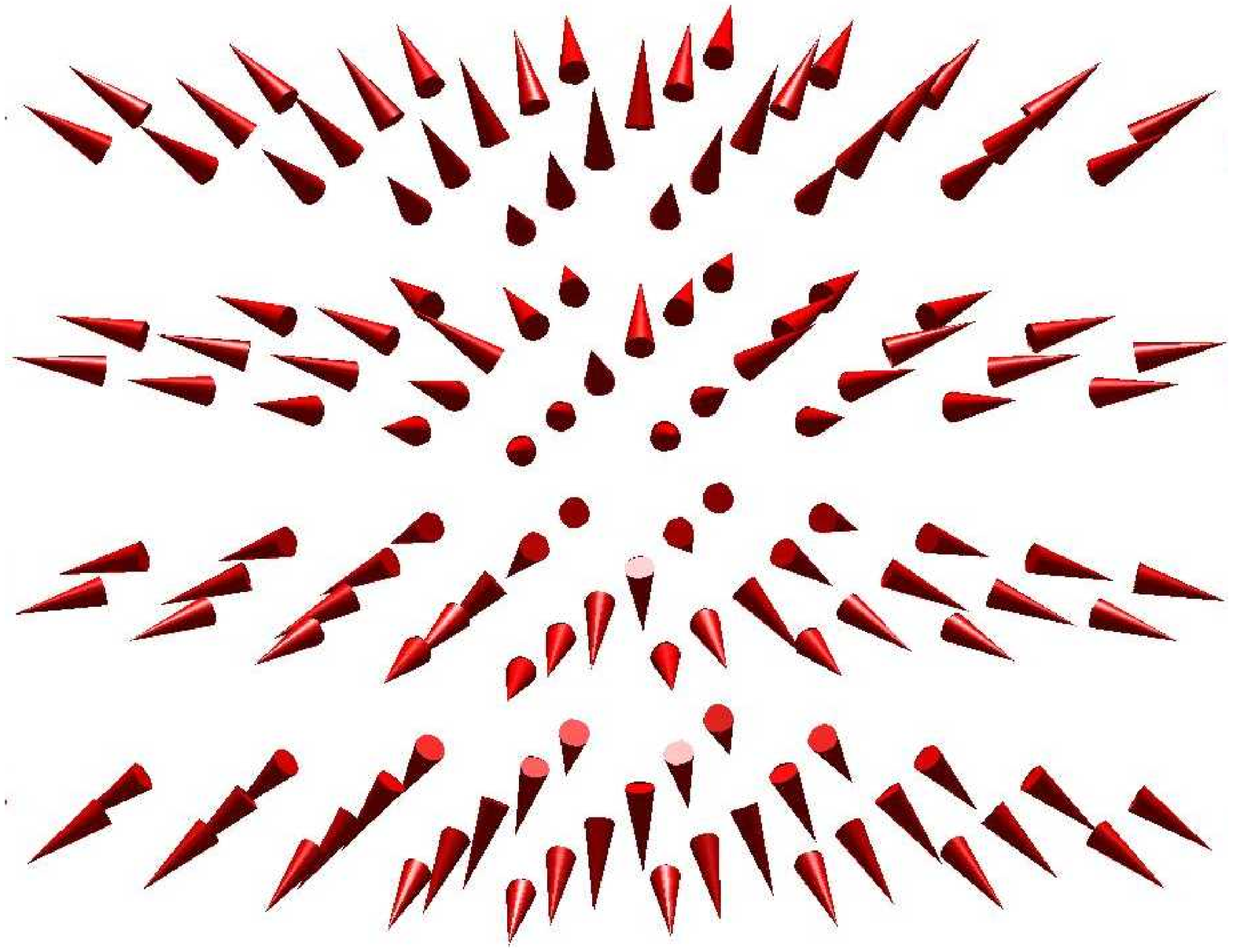} %
\includegraphics[width=0.45\columnwidth]{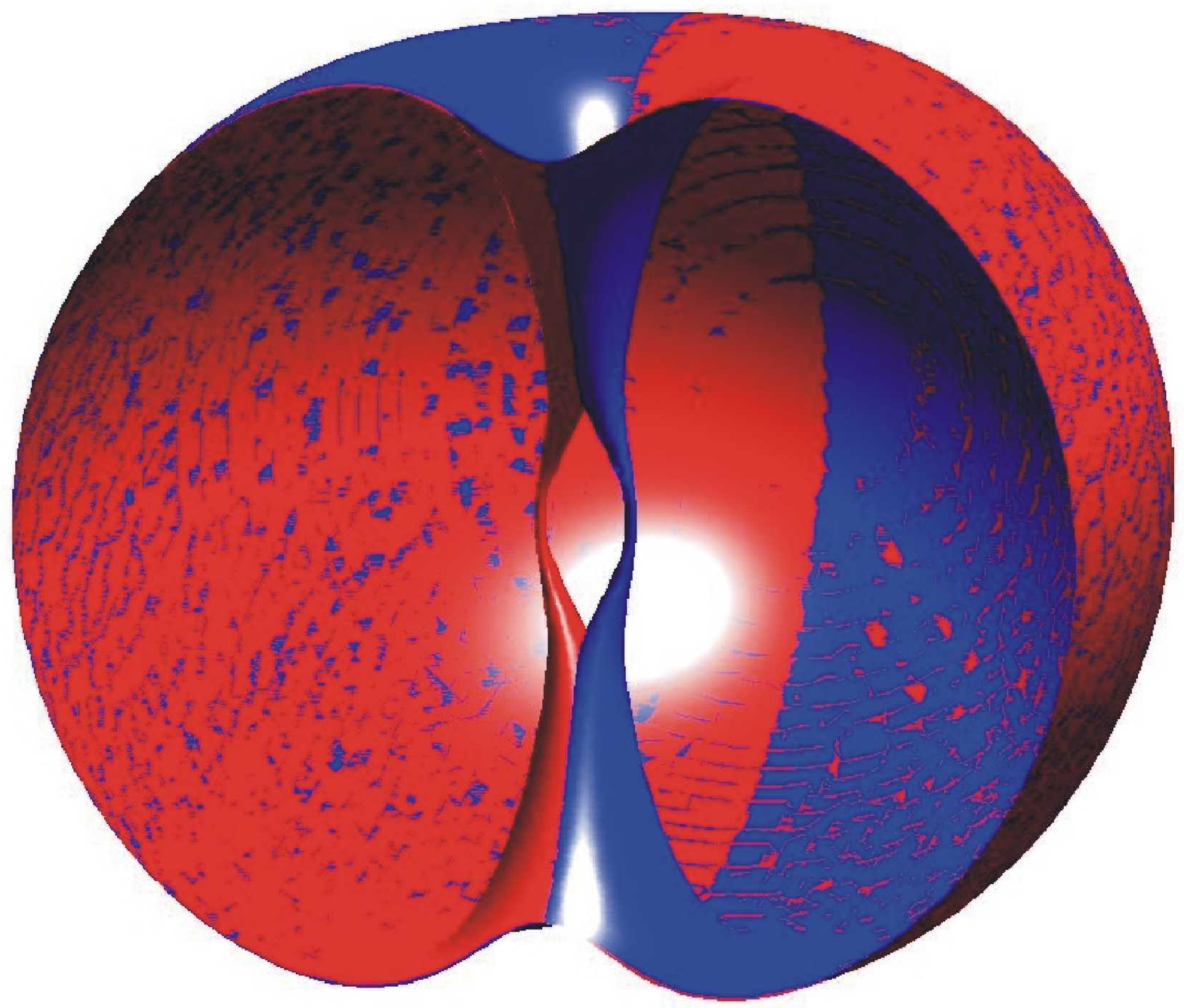} %
\caption{The stable half-quantum vortex ring (Alice ring), when the energy of an initial spherically symmetric monopole was minimized by continuously deforming the field configuration. The asymptotic distribution of the spin quantization axis $\dv(\rv)$ (left), for $r\gg \protect\xi _{a}$, forms the radial hedgehog. For visualization purposes, the unoriented $\dv(\rv)$ field is drawn by cones. We show the constant surface density plots (right) for $|\protect\psi _{1}(\mathbf{r})|^{2}$ (red) and for $|\protect\psi _{-1}(\mathbf{r})|^{2}$ (blue), where the monopole core is deformed for $r\alt\xi _{a}$ with the two line vortices separating.}
\label{ches}
\end{figure}

\begin{figure}[tbp]
\includegraphics[width=0.57\columnwidth]{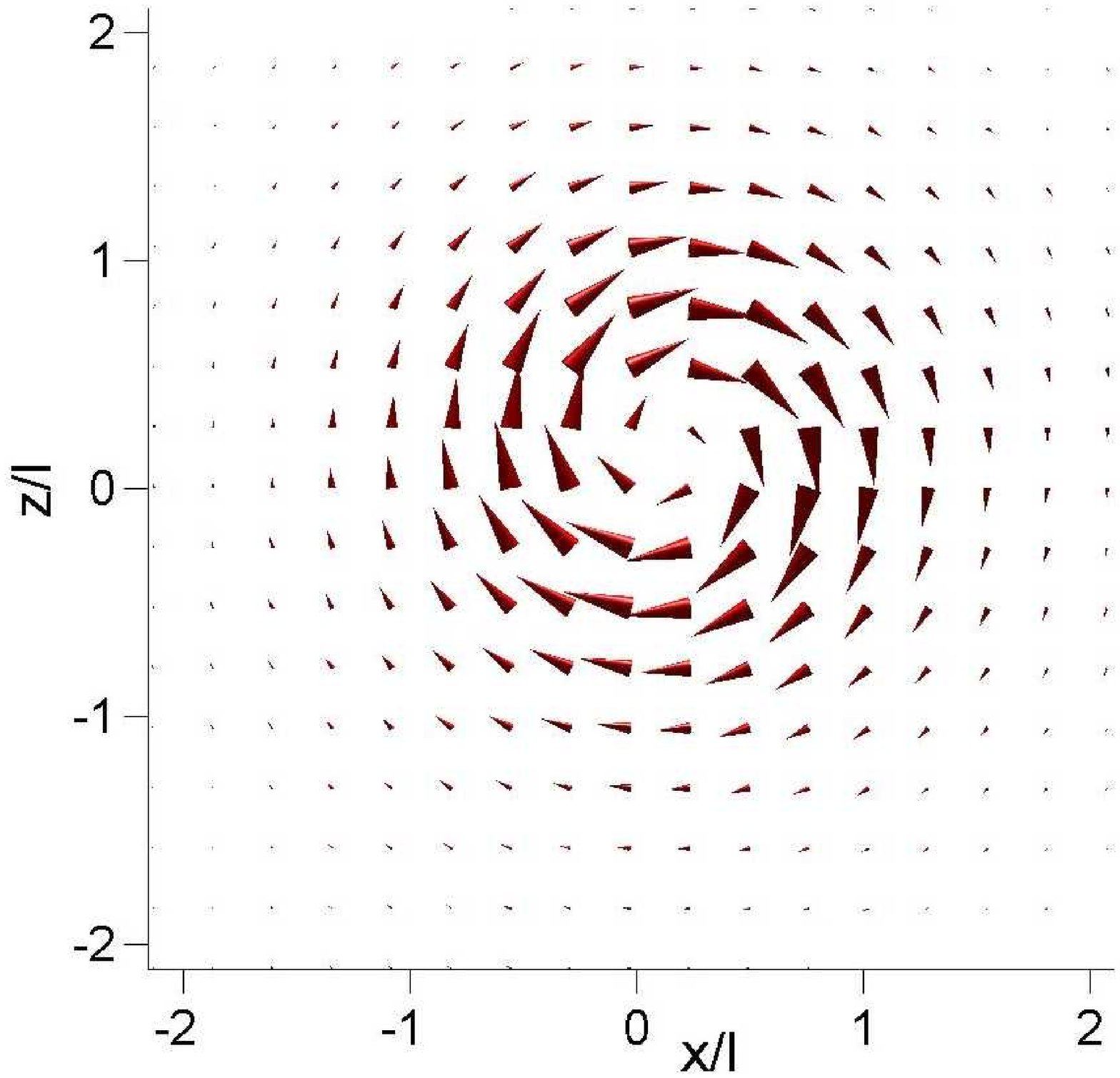} %
\includegraphics[width=0.37\columnwidth]{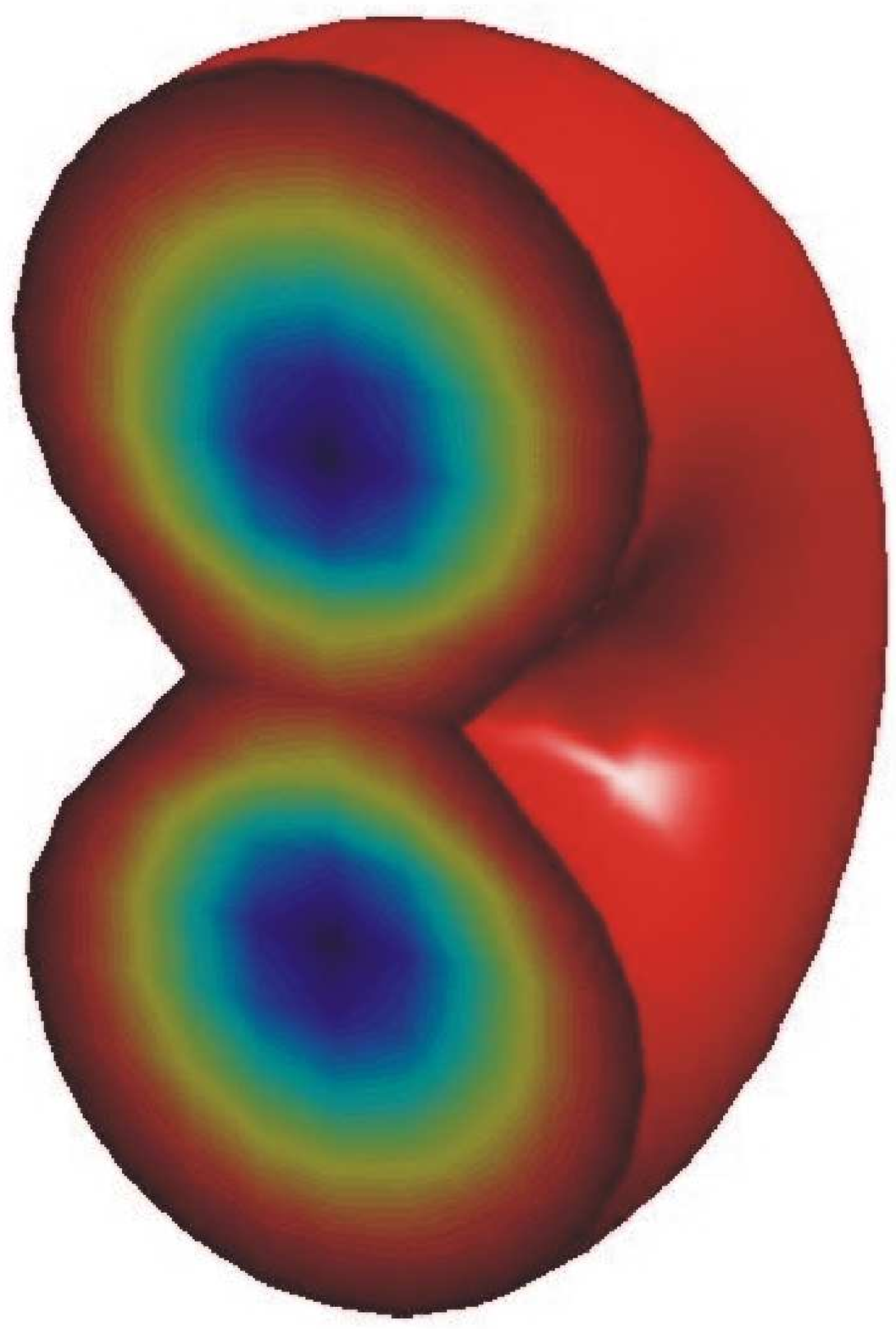}
\caption{The spin profile of the Alice ring displayed in Fig. \ref{ches}. The spin expectation value $\langle \mathbf{F}\>$ (left) is nonvanishing along the half-quantum vortex ring core. The absolute value of the spin $|\langle \mathbf{F}\>|^{2}$ (right) between the isosurface sections is indicated by a colormap from red ($|\langle \mathbf{F}\>|^{2}=0.35$) to purple ($|\langle \mathbf{F}\>|^{2}=1$). To display the isocaps of $|\langle \mathbf{F}\>|^{2}$ we have rotated the surface with respect to the spin field graph $\langle \mathbf{F}\>$ approximately $\pi/4$ counter-clockwise along the $z$ axis and cut the ring half.}
\label{ches2}
\end{figure}

We examine the recently presented case \cite{STO01} of a defect analogous to the 't Hooft-Polyakov monopole \cite{HOO74}, in a antiferromagnetic, or polar, spin-1 BEC \cite{LEA03}. We will show that it is only in the strongly antiferromagnetic regime, which is not attained in current experiments, that its stable core will be the spherically symmetric hedgehog, with a total density depression, of Ref.~\cite{STO01}. In the weakly antiferromagnetic regime that currently holds, the wavelength at which the antiferromagnetic constraint may be violated is much larger than that at which the total density constraint fails. The stable defect core therefore extends to this larger size, and holds non-zero average spin instead of a density zero. The singular point defect itself deforms to a circle: a half-quantum vortex ring (Figs.~\ref{ches} and~\ref{ches2}), called an `Alice ring' by high energy physicists \cite{SCH82,BAI95,STR03}, which carries a topological charge similar to delocalized magnetic `Cheshire' charge \cite{ALF91}. This forms an interesting connection between ultra-cold atom experiments and elementary particle physics. It also shows that dissipation, often an obstacle in state engineering, can sometimes perform the intricate final step in producing an exotic object.

We consider the BEC of spin-1 atoms. In the absence of a magnetic trapping potential, the macroscopic BEC wave function is determined by a spinor wave function $\Psi $ with three complex components \cite{HO98,Pethick}. The Hamiltonian density of the classical Gross-Pitaevskii (GP) mean-field theory for this system reads:
\beq \label{hamiltonian}
{\cal H} =-\frac{\hbar^2}{2m} |\nabla \Psi|^2
+ V \rho + {c_0\over2} \rho^2 + {c_2\rho^2\over2} |\< {\bf F}\> |^2 \, ,
\eeq
where ${\bf F}$ are the 3-by-3 Pauli spin matrices, $\< {\bf F} \>= \Psi ^{\dagger}\cdot {\bf F}\cdot \Psi/\rho$ denotes the average spin, $\rho=|\Psi|^2$ the total atom density, $c_0\equiv 4\pi\hbar^2(2a_2+a_0)/3m$, and $c_2\equiv 4\pi\hbar^2(a_2-a_0)/3m$, for $a_F$ the two-body $s$-wave scattering length in the total spin $F$ channel \cite{HO98}. For $^{23}$Na, $a_a\equiv (a_2-a_0)/3\simeq 2a_B$ and $a_s\equiv (2a_2+a_0)/3\simeq 50a_B$, indicating $c_2/c_0\simeq 0.04$, where $a_B = 0.0529$nm is the Bohr radius. Here $V$ denotes the external potential, for an isotropic optical dipole trap with the frequency $\omega$: $V(\rv)=m\omega^2 r^2/2$. For $^{23}$Na $c_2>0$, and the energy is minimized by setting $\langle {\bf F} \rangle ={\bf 0}$ throughout the BEC for the case of a uniform order parameter field. The zero average spin corresponds to the ground polar state, where we may determine all the degenerate states by means of the macroscopic BEC phase $\varphi$ and a real unit vector $\dv(\rv)$ defining the quantization axis of the spin. The BEC wave function then reads:
\beq
\Psi =
\begin{pmatrix}
\psi _{1} \\
\psi _{0} \\
\psi _{-1}
\end{pmatrix}
={\sqrt{\rho }\,e^{i\varphi }\over \sqrt{2}}
\begin{pmatrix}
-d_{x}+id_{y} \\
\sqrt{2} d_{z} \\
d_{x}+id_{y}
\end{pmatrix}\,.\label{polar}
\eeq
As in the similar polar phase of superfluid $^{3}$He-A
\cite{VOL90,VOL03,VOL77}, however, the states $(\mathbf{d},\varphi
)$ and $(-\mathbf{d},\varphi +\pi )$ are identical \cite{LEO00}.
The polar order parameter space, which may appear to be
$S^{1}\times S^{2}$, is actually factorized by the two-element
discrete group $Z_{2}$. Consequently, we take the $\mathbf{d}$
field to define unoriented axes rather than vectors.

\begin{figure}[tbp]
\includegraphics[width=0.95\columnwidth]{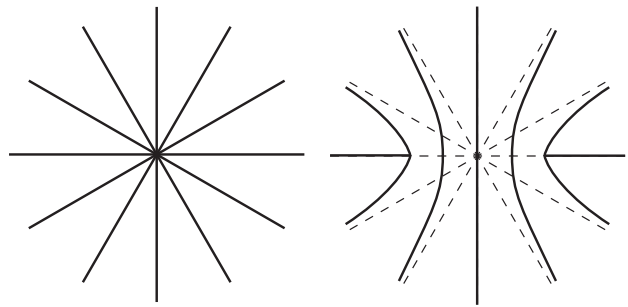}
\caption{Continuous deformation of the radial hedgehog into an Alice ring. Lines represent flow lines of the order parameter $\mathbf{d}$ field. We show planar sections of the monopole unperturbed (left) and deformed into an Alice ring (right). Note that the asymptotic behavior of $\dv$ remains unchanged.}
\label{defor}
\end{figure}

The spherically symmetric monopole $\Psi_M$ is obtained from
Eq.~(\ref{polar}) by the radial hedgehog field
$\dv(\rv)=\hat{\rv}\equiv (\sin \theta \cos \phi,\sin \theta \sin
\phi ,\cos \theta )$ \cite{BUS99}, with $\varphi=0$ and $\rho
=\rho _{M}(r)$ minimizing the energy of the symmetric
configuration \cite{STO01}. This is singular at the origin
indicating a point defect with $\rho(0)=0$. Spinor component
$\psi_0$ resembles a dark soliton and $\psi_{\pm 1}$ form
perfectly overlapping, straight singly-quantized vortex lines with
opposite circulation, perpendicular to the phase kink plane. The
topologically invariant winding number,
\begin{equation}\label{W}
W={\frac{1}{8\pi }}\int_{\partial \Omega }dS_{i}\,\epsilon _{ijk}\dv\cdot {\frac{\partial \dv}{\partial x_{j}}}\times {\frac{\partial\dv}{\partial x_{k}}}\,,
\end{equation}
is defined on any closed surface $\partial \Omega $ that encloses
the origin. Because the sign of $\mathbf{d}$ is ambiguous, though,
the sign of $W$ is arbitrary. Moreover, because the $\mathbf{d}$
field is actually unoriented, the monopole point defect may be
continuously deformed into a circular line defect, without
changing $W$ (on any surface enclosing the singular ring), by
punching a hole in the spherically symmetric core; see
Fig.~\ref{defor}. To keep $\Psi$ single-valued on the disc bounded
by the ring, the macroscopic phase $\varphi$ must change by $\pi$
around any loop that links the defect circle, while there is also
a $\pi$-disclination in $\dv$ on the disc. We identify this
structure as a half-quantum vortex line \cite{VOL90,VOL03},
forming a closed circular ring, also called an Alice ring
\cite{SCH82,BAI95,STR03,ALF91}. To keep $\Psi $ single-valued on
the ring itself, one can either have $\rho$ vanish there, or have
$|\< {\bf F}\> |= 1$ on the ring instead.

Thus, because of the ambiguity of the direction of $\dv$ in the
polar state, different core structures can be smoothly deformed
into each other. The core stability is therefore solely determined
by energetic considerations. Physically, the $\rho =0$ ring
evidently has higher energy than the $\rho =0$ point; but
energetic stability of the point defect against the Alice ring
with $\left| \langle \mathbf{F}\rangle \right|=1 $ core is a
nontrivial question. We may define two healing lengths $\xi
_{s}\equiv (8\pi a_{s}\rho )^{-1/2}$ and $\xi_{a}\equiv (8\pi
a_{a}\rho )^{-1/2}$, in Eq.~(\ref{hamiltonian}). They describe the
length scales over which $\rho$ and $|\<{\bf F}\>|$, respectively,
tend to their bulk value when subjected to localized
perturbations. (We can also say that $\xi_s$ and $\xi_a$
correspond to the excitation wavelengths below which density and
spin fluctuations cease to be energetically suppressed.) Since at
the defect core we may have either $\rho=0$ or $|\<{\bf F}\>|= 1$,
$\xi_s$ and $\xi_a$ determine the core sizes in the two cases. We
numerically demonstrate, by decreasing the ratio $a_a/a_s$, that
the spherically symmetric monopole core with the total density
suppression becomes unstable to formation of an energetically
stable Alice ring with $|\<{\bf F}\>|= 1$ core. We find the radial
hedgehog to be unstable in a linear stability analysis in bulk at
$c_2/c_0\lesssim 0.17$ and in the mean-field theory in a trap at
$c_2/c_0\lesssim 0.16$, using an experimentally feasible set of
parameters for $^{23}$Na.

Before turning to the full nonlinear mean-field theory of the trapped BEC, we study the linear stability of the radial hedgehog in the homogeneous case. We expand the Hamiltonian (\ref{hamiltonian}), with $V=0$, to second order around $\Psi _{M}$. Perturbations of the form
\begin{equation}
\delta \Psi =i\frac{\eta (r)}{\sqrt{2}}
\begin{pmatrix}
-C_{1}+iC_{2} \\
\sqrt{2}C_{3} \\
C_{1}+iC_{2}
\end{pmatrix}
+i\frac{\mathbf{C}\cdot \mathbf{\hat{r}}}{\sqrt{2}}\,\varsigma (r)
\begin{pmatrix}
-e^{-i\phi }\sin \theta  \\
\sqrt{2}\cos \theta  \\
e^{i\phi }\sin \theta
\end{pmatrix}\label{deltaPsi}
\end{equation}
decouple from all others, for real $C_{j}$, $\eta $ and $\varsigma $, if $\eta $ and $\varsigma $ satisfy (here $X\equiv c_2/c_0$ and $\bar\rho\equiv \rho_M(r)/\rho_0$, where $\rho_0$ denotes the constant asymptotic value of the density)
\begin{align}
\lambda \varsigma  &=-{\frac{1}{2r^{2}}}\left( r^{2}\varsigma ^{\prime }\right)
^{\prime }+{\frac{3}{r^{2}}}\varsigma +[(\bar\rho-1)\varsigma -2{X}\bar\rho\eta]{1\over\xi_s^2}\nonumber \\
\lambda \eta  &=-{\frac{1}{2r^{2}}}\left( r^{2}\eta ^{\prime }\right)
^{\prime }+\left[ \bar\rho\left( 1+2{X}\right) -1\right] {\eta\over\xi_s^2} -\frac{\varsigma}{r^{2}}
\end{align}
for some real eigenvalue $\lambda$. For small $C=|\mathbf{C}|$,
the change in free energy due to this perturbation in $\Psi $ will
be $\delta E=\lambda C^{2}m/\hbar^2$. Numerical solution shows
\cite{US2} that there exists one negative $\lambda$, hence three
degenerate instabilities proportional to $C_i$, when $c_2/c_0\alt
0.17$. We can show that no other instabilities exist in bulk, so
for $c_2/c_0\agt 0.17$ the point defect is stable. Since the
perturbation (\ref{deltaPsi}) gives
\begin{equation}
\<{\bf F}\> =\frac{2\mathbf{C}\times \mathbf{\hat{r}}\sqrt{\rho _{M}}\eta }{\rho
_{M}+C^{2}\eta ^{2}},
\end{equation}
where $\<{\bf F}\>$ is with respect to the state $\Psi _{M}+\delta \Psi $, we have $\left| \langle \mathbf{F}\rangle \right| =1$ on a circle in the plane through the origin perpendicular to $\mathbf{C}$, at radius $r_{\ast }$, such that $\rho _{M}( r_{\ast }) =[C\eta ( r_{\ast })]^2$. The form of $\rho _{M}(r)$ (rising monotonically from 0 to $\rho_0$) and $\eta $ (decreasing monotonically to zero as $r\rightarrow \infty $) ensures that $r_{\ast }$ grows monotonically with $C$. As a result, a randomly oriented Alice ring will form spontaneously from the symmetric monopole, as $C$ grows from zero by relaxation, for $c_2/c_0\alt 0.17$.

In the full 3D classical mean-field theory of the spin-1 monopole in an isotropic trap, we minimized the energy by evolving the GP equations,
\beq \label{GP}
i \, \hbar \frac{\partial \Psi}{\partial t} = (-\frac{\hbar^2}{2m}{\bf \nabla}^2+V+ c_0 \rho)  \Psi
+ c_2 \rho \<{\bf F}\>\cdot  {\bf F}\cdot \Psi \, ,
\eeq
in imaginary time.
The initial state was an approximate spherically symmetric monopole solution with a point core, embedded in a Thomas-Fermi density profile. The integration was performed on a spatial grid of $128^3$ points using the split-step method. At every time step we normalized the wave function to fix the total atom number. The numerical simulations were fully 3D without imposing any symmetry on $\Psi $ as it relaxed.

We varied the relative strength of the two interaction coefficients $c_0$ and $c_2$, as well as the total atom number $N$. The dynamics depends on the dimensionless interaction strengths $c_0'=4\pi Na_{s}/l$ and $c_2'=4\pi Na_{a}/l$, where $l\equiv [\hbar /(m\omega )]^{1/2}$. Hence, for a given atom with fixed values of scattering lengths, the results are unchanged for any scaling of length and time, which does not change $N^2\omega$. In optical dipole trap experiments on $^{23}$Na BECs, a wide range of aspect ratios and trapping strengths have been demonstrated, with the typical values of $l/a_s$ varying between 500 and 4000 \cite{LEA03}.

For small values of $c_2/c_0$, the point core deformed into a half-quantum vortex ring. With the initial radial hedgehog at the trap center we found vortex ring configurations as local energetic minima, with the size of the ring approximately determined by $\xi_a$\cite{comment}. In Figs.~\ref{ches} and~\ref{ches2} we display such results using the parameters of the spin-1 $^{23}$Na with $c_2/c_0=0.04$ and $c_0'=2\times10^4$. With the trapping frequency $\omega=2\pi\times10$Hz, this corresponds to $N\simeq4\times10^6$ atoms. In the initial state of the radial hedgehog, the components $\psi _{\pm 1}$ form two oppositely circulating vortex lines with perfectly overlapping density profiles. Due to dissipation, the two vortices in $\psi _{\pm 1}$ separated near the trap center, forming an Alice ring. Note that this corresponds to the growth of the perturbation $C_{2}$ in Eq.~(\ref{deltaPsi}). The numerical noise in this example effectively seeded the instability $C_2$ yielding
\begin{equation}
\label{C2}
\psi^{(M)}_{\pm 1}+\delta \psi _{\pm 1}\varpropto \frac{\sqrt{\rho _{M}}}{r}\left[ \left(\mp x - \frac{C_{2}r\eta }{\sqrt{\rho _{M}}}\right) + i  y \right].
\end{equation}
Since $r\eta /\sqrt{\rho _{M}}$ is constant at the origin, but vanishes as $r\rightarrow \infty$, this means that near the center the two vortices in $\psi _{\pm 1}$ separate in the $xz$ plane. Growth of $C_{1}$ involves a similar separation in the $yz$ plane. One can clearly see the $\pi$ winding of the macroscopic phase around closed loops threading the Alice ring in Eq.~(\ref{C2}).

With some initial parameters we also observed the radial hedgehog evolving to an Alice ring with the vortex cores in $\psi_{\pm1}$ expanding, but not separating, while the phase kink in $\psi_0$ no longer maintained a total density depression at the center. This clearly represents the growth of the perturbation $C_3$ in Eq.~(\ref{deltaPsi}), since the term proportional to $\varsigma \left(r\right)$ is simply a perturbation in the macroscopic phase $\varphi$ of Eq.~(\ref{polar}). In the general case of a symmetric initial monopole, dissipation spontaneously breaks the spherical symmetry and the resulting ring has an arbitrary orientation.

It is easy to see that the Alice ring in bulk must stabilize at some finite value of the radius, because the amount of energetically costly non-zero $\langle \mathbf{F}\rangle $ rises as the ring grows. This was also true in a trap for small enough rings. However, with smaller atom numbers, $c'_0\alt 10^4$ for $^{23}$Na, a larger ring resulted, which was destabilized by the inhomogeneous density profile: the two vortex lines in $\psi _{\pm1}$ completely detached and left the atomic cloud in opposite directions.

The half-quantum vortex ring, shown in Figs.~\ref{ches} and~\ref{ches2}, was still unstable with respect to drifting out of the atomic cloud as a unit if initially displaced from the trap center. This is because of the reduced order parameter bending energy at lower atom densities. However, we successfully stabilized the Alice ring by creating a local density minimum at the trap center, by adding to the harmonic trapping potential the optical potential $V_L$, simulating two orthogonal blue-detuned focused Gaussian laser beams: $V_L= V_1\exp{[-2(x^2+y^2)/w^2]}+V_2\exp{[-2(x^2+z^2)/w^2]}$.
We note that the Alice ring can always be made long-living on experimental time scales even without the stabilizing field, since the drift time should increase with the BEC size, while the ring formation time should decrease.

We also investigated the onset of the point defect core instability by varying $c_2'$ for a given $c_0'=2\times10^4$. We found the symmetric monopole to be unstable for $c_2/c_0\lesssim0.16$, in a good agreement with the linear stability analysis in bulk. The precise value of the threshold is very difficult to determine in a trap, since the convergence in the simulations is particularly slow close to the threshold. In the case of very long runs, numerical noise also eventually displaces the monopole so that it starts drifting out of the atomic cloud. For smaller values of $c_2'$, representing unstable Alice rings which break apart due to the inhomogeneous density, the instability of the point defect occurred with much larger values of $c_2/c_0$, when the trap length scale became comparable to the polar healing length $l\sim \xi_a$, emphasizing the nontrivial nature of the finite size effects.

In the experimental preparation of an Alice ring by relaxation, it is helpful to note that defects and textures in multi-component BECs may generally be viewed as combinations of vortex lines, rings, and phase kinks \cite{RUO01}. As we already noted, the radial hedgehog has an especially simple structure of vortex lines and phase kinks which have been experimentally created in individual atomic BEC components, e.g., by means of localized laser fields, which are rapidly swept around the trap during a Raman transition \cite{MAT99,AND01,RUO00}. The spherically symmetric monopole could similarly be created, by a sequence of Raman pulses, in a straightforward generalization of the pulse sequences proposed in Ref.~\cite{RUO01}. Dissipation plays a crucial role in the state engineering process and will then perform the final step of generating a stable Alice ring by spontaneously breaking the spherical symmetry. The stable radius $\xi_a$ of the Alice ring may easily extend over several microns, making the core quite observable, possibly even without ballistic expansion.

We can summarize our results in basic terms. In $^{23}$Na BECs the
weakness of the antiferromagnetic energy ($a_{s}\gg a_{a}$),
topology, and the gradient energy of the order parameter, may
conspire to favor nonvanishing spin values, even for $a_{a}>0$: It
is energetically more favorable to violate the antiferromagnetic
constraint than to force the superfluid density to vanish. As a
result, a monopole core deforms to a ring and exhibits a
nonvanishing spin expectation value and a non-zero superfluid
density. In other words, the strong order parameter bending energy
close to the singular defect mixes the polar and the ferromagnetic
phases of the spinor BEC, rather than forcing the total superfluid
density to zero at the singularity. Due to the length scale
hierarchy, the stable size of the defect core will then be
determined by $\xi _{a}$ instead of the much smaller $\xi _{s}$.
And the creation of this intricate Alice ring structure will occur
spontaneously, from the much simpler symmetric monopole, by
relaxation alone.

While the motion and interaction of point defects was analyzed in
\cite{STO01}, the dynamics of Alice rings remains a question for
future theoretical and experimental study.  So does the possible
meta-stability of Alice rings for $c_2\agt 0.17 c_0$. Topological
defects in spinor BECs promise a rich phenomenology, which current
experimental techniques will allow us to explore.

\acknowledgments
{This research was supported by the EPSRC.}


\begin{thebibliography}{99}
\bibitem{comment} The conservation of the topological charge in the core deformation was confirmed by numerically integrating the winding number (\ref{W}) during the evolution of the point defect to a ring, yielding $W\simeq 1$ within a couple of percents, when the surface was sufficiently large to enclose the whole ring.
\end{thebibliography}
\end{document}